\documentclass[twocolumn,prl,aps]{revtex4}
\usepackage{amsmath}
\usepackage{amssymb}
\usepackage{graphicx}
\usepackage{dcolumn}
\usepackage{bm}
\usepackage{xcolor}
\usepackage{epstopdf}

\begin{document}

\preprint{Phys.Rev.Lett.}

\title{Two-dimensional topological insulator state in double HgTe quantum well}

\author{G. M. Gusev,$^1$ E. B. Olshanetsky,$^{2}$ F. G. G. Hernandez,$^1$
O. E. Raichev,$^3$ N. N. Mikhailov,$^2$ and S. A. Dvoretsky$^2$ }

\affiliation{$^1$Instituto de F\'{\i}sica da Universidade de S\~ao
Paulo, 135960-170, S\~ao Paulo, SP, Brazil}
\affiliation{$^2$Institute of Semiconductor Physics, Novosibirsk
630090, Russia}
\affiliation{$^3$Institute of Semiconductor Physics, NAS of
Ukraine, Prospekt Nauki 41, 03028 Kyiv, Ukraine}

\date{\today}

\begin{abstract}

The two-dimensional topological insulator phase has been observed previously in single
HgTe-based quantum wells with inverted subband ordering. In double quantum wells (DQWs),
coupling between the layers introduces additional degrees of freedom leading to a rich
phase picture. By studying local and nonlocal resistance in HgTe-based DQWs, we observe
both the gapless semimetal phase and the topological insulator phase, depending on
parameters of the samples and according to theoretical predictions. Our work establishes
the DQWs as a promising platform for realization of multilayer topological insulators.


\end{abstract}

\maketitle
Topological insulators (TIs) are a new class of materials characterized by an insulating
band gap in the bulk, like in ordinary band insulators, but having topologically protected
conducting states at their edges for two-dimensional (2D) TIs or at the surface for three-dimensional
TIs \cite{Kane, Bernevig, Hasan, Qi}. The TI phase has been theoretically predicted \cite{Bernevig}
and experimentally realized in HgTe quantum wells (QWs) with Cd$_x$Hg$_{1-x}$Te barriers \cite{Konig}.
The crucial properties enabling the 2D TI states in these systems are the strong spin-orbit
coupling that leads to inverted band structure of HgTe and the size quantization that opens
the gaps in the electron spectrum of initially gapless HgTe. In HgTe-based QWs with the well width
$d$ above the critical one, $d > d_{c}$, the upper size-quantization subband of heavy holes stays above
the electronlike subbands. This inverted subband ordering produces the 2D TI phase with
one-dimensional edge states inside the gap in the bulk 2D subband spectrum. Electron transport
in the edge states is expected to be highly ballistic because the elastic
backscattering of electrons is forbidden due to the time reversal symmetry in the absence
of magnetic field and magnetic impurities. Experimental manifestations of the edge transport
are the quantization of conductance at the universal value $2e^{2}/h$ and the appearance of
nonlocal voltage away from the dissipative bulk current path \cite{Konig,Roth}.

\begin{figure}[ht!]
\includegraphics[width=9cm,clip=]{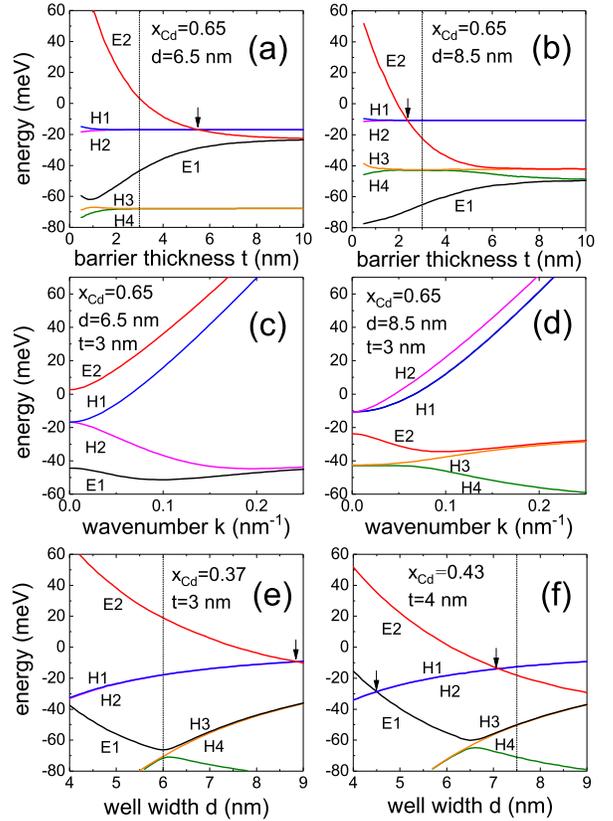}
\caption{\label{fig.1}(Color online) Calculated band structure for symmetric [013]-grown
HgTe/Cd$_{x}$Hg$_{1-x}$Te DQWs with different well widths $d$, barrier thicknesses $t$, and
Cd content in the barriers. (a,b): dependence of subband energies at $k = 0$ on $t$,
(c,d): subband dispersion for ${\bf k} || [100]$, (e,f): dependence of subband energies
at $k = 0$ on $d$. The arrows show the points of subband inversion leading to phase transitions,
and the vertical lines correspond to parameters of the DQWs studied in our experiment.}
\end{figure}

Double quantum wells (DQWs) are bilayer systems consisting of two QWs separated by
a tunneling-transparent barrier. Theoretical studies suggest \cite{kristopenko} that the picture
of phase states in these structures is strongly affected compared to the single QW case and becomes
more rich because of the additional degrees of freedom associated with an increase in the number
of 2D subbands and tunnel-induced hybridization between them. In particular, the subband ordering
depends not only on the well width $d$ and Cd content in the barriers $x_{Cd}$ but also
on the barrier thickness $t$, and is sensitive to the asymmetry induced, for example,
by a transverse electric field. Figure 1 (a,b) shows a typical
dependence of subband edges on $t$ at a fixed $d$ for symmetric HgTe-based wells (see \cite{suppl}
for the details of calculations). The electronlike subbands E1 and E2, formed
as a result of coupling between the $\Gamma_6$ states and the light hole part of $\Gamma_8$
states, are characterized by the wave functions which strongly penetrate into the
barriers and, therefore, show a significant tunnel-induced splitting. In
contrast, the hole subbands H1-H4, originating from the heavy hole part of $\Gamma_8$ states,
are characterized by the wave functions weakly penetrating into the barriers because of the
large effective mass. Thus, the tunnel-induced splitting of the pairs of subbands H1-H2 and
H3-H4 exists only at very small $t$, while for $t > 2$ nm their energies are the same as
in the single QWs. The degeneracy of H1 and H2 subbands is lifted at nonzero wave vector
${\bf k}$, forming two branches with the opposite dispersion in the case of mixed ordering
of subbands, E1-H2-H1-E2, which results in a gapless energy spectrum, Fig. 1 (c).
With increasing $t$ and/or $d$, a fully inverted ordering, E1-E2-H2-H1, is realized,
and a gapped TI phase appears, Fig. 1 (d). The energy spectrum in the gapless phase
resembles the bilayer graphene spectrum \cite{castro1} with the absence of valley
degeneracy \cite{note}.

Similar as in bilayer graphene \cite{castro2}, it is predicted that the gapless phase in symmetric
HgTe DQWs can be transformed into a gapped phase by breaking the inversion symmetry of two wells
via a transverse electric field due to asymmetric doping or external gate biasing
\cite{michetti,kristopenko}. Since the gap energy depends on the field, both these
systems are the tunable gap semiconductors. However, there is a principal difference
between the biased bilayer graphene and asymmetric HgTe DQWs: in the latter case the topologically
protected edge states appear in the gap. Indeed, in contrast to bilayer graphene, HgTe-based DQWs
possess topological properties due to subband inversion, and the calculations based on the effective
Hamiltonian models \cite{michetti,kristopenko,baum} show coexistence of the bulk and edge states
in the gapless semimetal phase. Thus, it is expected that opening the gap in HgTe DQWs by the field
makes them 2D TIs, where the edge state transport can be detected experimentally.

In this Letter, we present the results of studies of the local and nonlocal resistance in
HgTe-based DQWs. By studying several groups of samples with different parameters $x_{Cd}$,
$d$, and $t$, we have observed both the gapped phase and the gapless semimetal phase, clearly distinguished by the temperature dependence of the local resistance. By measuring nonlocal
resistance in the gapped samples, we confirm the edge state transport in the absence of
magnetic field, thereby providing the experimental evidence of the 2D TI phase in HgTe DQWs.
Furthermore, we observe the gap and the edge state transport in the DQWs which are expected
to be in the gapless phase if the transverse electric field were absent.

Double quantum wells HgTe/Cd$_{x}$Hg$_{1-x}$Te with [013] surface orientation and equal
well widths from 6.0 to 8.5 nm were prepared by molecular beam epitaxy (MBE) \cite{suppl}.
The layer thicknesses were determined by ellipsometry during the MBE growth, with the
accuracy $\pm 0.3$ nm. The devices are designed for multiterminal
measurements and contain three 3.2 $\mu$m wide consecutive segments of different
length (2, 8, and 32 $\mu$m) and 9 contacts (Fig. 2) shunting both QW layers. A dielectric
layer, 200 nm of SiO$_{2}$, was deposited on the sample surface and then covered by a TiAu gate.
The density variation with the gate voltage was $0.86 \times 10^{11}$ cm$^{-2}$ V$^{-1}$. The
transport measurements were performed in the range of temperatures $T$ from 1.4 K to 70 K by
using a standard four-point circuit with 1-13 Hz ac current of 1-10 nA through the sample,
which is sufficiently low to avoid overheating effects. For each set of parameters, results
for two typical devices (1,2) are represented. Table I lists parameters of the devices,
including the gate voltage corresponding to the charge neutrality point (CNP) at $T=4.2$ K.

\begin{table}[ht]
\caption{\label{tab1} Parameters of HgTe/Cd$_{x}$Hg$_{1-x}$Te DQWs.}
\begin{ruledtabular}
\begin{tabular}{lccccc}
  $d$ (nm) & $t$ (nm) & $x_{Cd}$ & $V_{CNP}$ (V) & properties \\
\hline
 8.5 & 3& 0.65 & -0.7  & TI \\
\hline
 7.5 & 4 & 0.43 & -0.8 &  narrow-gap TI  \\
\hline
 6.5 & 3 & 0.65 & -5 & gapless  \\
\hline
 6.0 & 3& 0.37 &-2.4 & TI \\

\end{tabular}
\end{ruledtabular}
\end{table}

\begin{figure}[ht!]
\includegraphics[width=9cm]{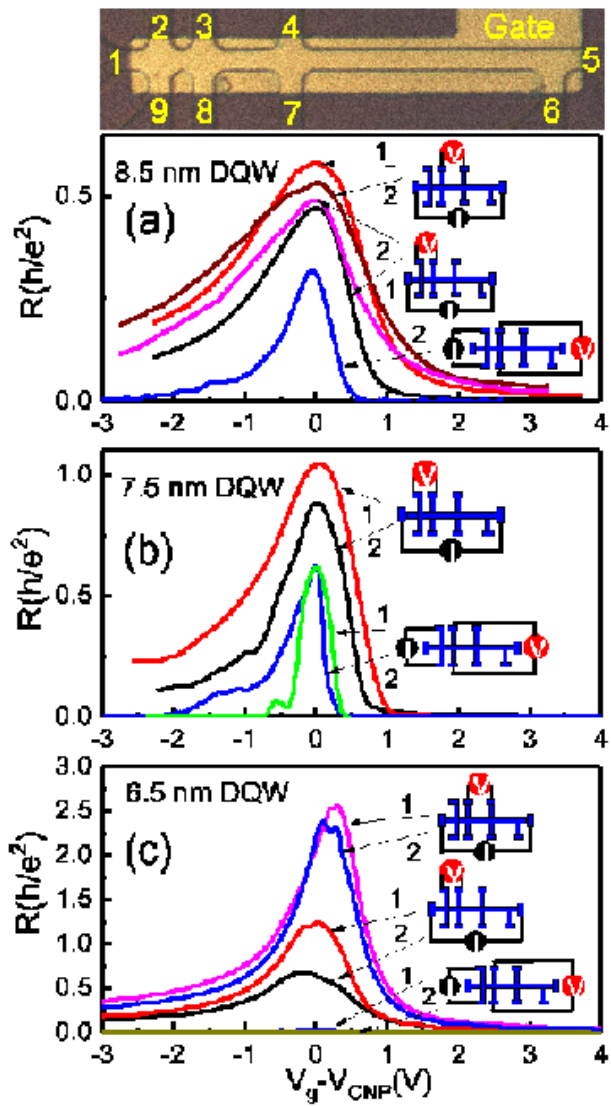}
\caption{(Color online)
Resistance at $T=4.2$ K as a function of the gate voltage $V_g$ for different configurations of
measurements and for different devices (1,2) fabricated from DQW structures with $d=8.5$ nm (a),
$d=7.5$ nm (b), and $d=6.5$ nm (c). The top panel shows the device layout and numbering
of the probes.}
\end{figure}

Figure 2 presents the local and nonlocal resistances at zero magnetic field for the
samples fabricated from the structures with $d=8.5$ nm, $d=7.5$ nm, and $d=6.5$ nm.
In the local configuration, the current $I$ flows between the contacts 1 and 5, and the voltage $V$
is measured between the short distance separated probes 2 and 3, $R_{L}=R^{2,3}_{1,5}=V_{2,3}/
I_{1,5}$, and between widely separated probes 3 and 4, $R_{L}=R^{3,4}_{1,5}=V_{3,4}/I_{1,5}$.
In the nonlocal configuration, the current flows between the contacts 2 and 9 while the voltage
is measured between the probes 3 and 8, $R_{NL}=R^{3,8}_{2,9}=V_{3,8}/I_{2,9}$. A slower decrease
of the resistances with $V_g$ in the negative voltage region is related to a larger density of
states in the valence band.

We first focus on the results obtained in DQWs with $d=8.5$ nm and 7.5 nm. In the first
one, the peaks of the resistances $R^{2,3}_{1,5}$ and $R^{3,4}_{1,5}$
are close to the quantity $5/9 (h/e^{2})$ expected in the case of purely ballistic edge state
transport in the multiprobe device shown in Fig. 2, though $R^{3,4}_{1,5}$ is larger
than $R^{2,3}_{1,5}$. In the second one, the peak resistance $R^{2,3}_{1,5}$ is considerably
larger than $5/9 (h/e^{2})$. These observations contradict with the picture of ballistic
edge state transport in 2D TIs implying that the backscattering is forbidden at sufficiently
low temperatures \cite{Kane, Bernevig, Hasan, Qi} so the resistance is independent of the distance between the voltage probes.
While some experimental works indeed confirm the quantized transport
in multiprobe HgTe QW devices \cite{Roth, Olshanetsky}, the longest edge channel length
that shows a quantized resistance peak never exceeds 10 $\mu$m. We notice that the edge
current in our experiment flows along the gated sample edge (see the top of Fig. 2), so
the effective length $L_{nm}$ of the edge channel segments between the probes $n$ and
$m$ is longer than the geometrical distance (bulk current path) between these probes and
is estimated as $L_{23}\simeq 6$ $\mu$m and $L_{34}\simeq 12$ $\mu$m \cite{suppl}. Thus,
the absence of ballisticity is expectable.

The lack of robustness of the ballistic edge state transport has attracted a lot of attention,
and numerous explanations have been proposed in recent years \cite{maciejko}. A detailed discussion
of the existing theoretical models is beyond the scope of our paper. We emphasize
that the edge state transport persists in both ballistic and diffusive regimes, while the
values of local and nonlocal resistances depend on the sample design and measurement
configurations. The most important characteristic feature of the 2D TIs, indicating the presence
of the edge state transport, is the existence of nonlocal resistance of the order of resistance
quantum $h/e^{2}$ \cite{Roth}.
We observe such nonlocal resistances in 8.5 nm and 7.5 nm DQWs. Figure 2 (a) and (b) shows a
significant difference between $R_{L}$ and $R_{NL}$: the peaks of $R_{NL}$ are narrower and this resistance disappears outside the bulk gap, where the edge channels are short circuited by the
bulk conduction. Similar results were obtained for other nonlocal measurement configurations.

The Landauer-B\"uttiker formalism \cite{Buttiker} allows one to find the resistance for
any measurement configuration in multiprobe samples, by taking into account a finite
probability of backscattering described by the mean free path length $l$ \cite{suppl}:
\begin{equation}
R^{i,j}_{n,m} \equiv \frac{V_{i,j}}{I_{n,m}} =\frac{h}{e^2}
\frac{(N_{ij}+L_{ij}/l)(N_{nm}+L_{nm}/l)}{N+L/l}.
\end{equation}
Here, $N_{ij}$ and $L_{ij}$ (or $N_{nm}$ and $L_{nm}$) are the number of edge channel segments
and their total length between the voltage (or current) leads along the arc that does not contain
the current (or voltage) leads. Next, $N$ is the total number of segments, which is equal
to the total number of leads, and $L$ is the total length of all segments, i.e., the perimeter
of the sample. In the ballistic transport limit, $L_{ij}/l \ll N_{ij}$, the resistance is an
integer fraction of $h/e^2$ \cite{Roth, Olshanetsky}, while in the general case the resistance
is sensitive to $L_{ij}$ and is larger than the ballistic one. However, a comparison of Eq. (1)
to the experimental peak resistances does not show a general agreement. The deviations are
strong for the nonlocal configuration, where the ballistic limit corresponds to
$R^{3,8}_{2,9}=10/9 (h/e^{2})$ while the measured resistances are not
larger than this value, as expected from Eq. (1) \cite{suppl}, but, in fact, are considerably
smaller. This means that the backscattering within the edge channel alone cannot explain the
transport, and the other processes responsible for attenuation of the voltage signal, such as
edge to bulk leakage of electrons and bulk current contribution, are important. Nevertheless,
the very existence of nonlocal resistance is a sufficient proof for the domination of the edge
transport near CNP in the samples with $d=8.5$ nm and $d=7.5$ nm, which also suggests the
presence of the bulk gaps.

Now we turn to the observations for DQW structure with $d=6.5$ nm, Fig. 2 (c), which differs
from the structure of Fig. 2 (a) only by the well width. One can see that the ratio $R^{3,4}_{1,5}/R^{2,3}_{1,5}$ is close to the ratio of the distances between the probes, which
suggests the dominance of the bulk transport. More important, the nonlocal resistance is absent.
This is the property associated with zero gap spectrum, shown in Fig. 1 (c) for this structure.
Unlike the edge state conductivity in 8.5 nm and 7.5 nm DQWs, the conductivity in 6.5 nm devices
is due to the bulk, while the edge contribution is short circuited.

\begin{figure}[ht!]
\includegraphics[width=8.5cm]{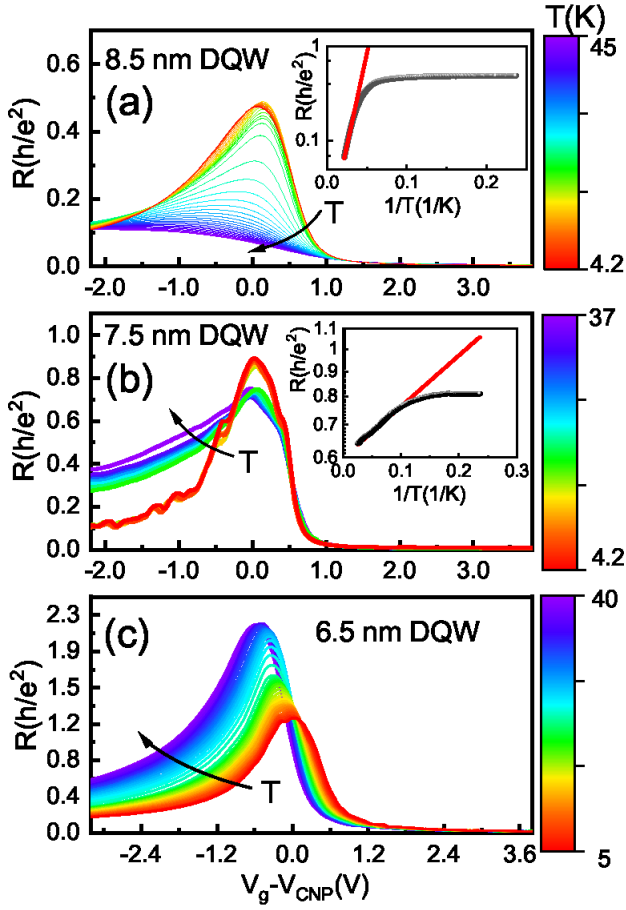}
\caption{(Color online)
Resistance as a function of the gate voltage for different temperatures for the DQWs with $d=8.5$ nm,
device 1 (a), $d=7.5$ nm, sample 2 (b), and $d=6.5$ nm, sample 1 (c). The insets show the
resistance at CNP as a function of $1/T$, the red lines correspond to $R \sim \exp(\Delta/2
k_B T)$ with $\Delta \simeq 15$ meV (a) and $\Delta \simeq 0.5$ meV (b).}
\end{figure}

For a further proof of the different nature of the transport in the samples with open and closed
bulk gap, we have measured the temperature dependence of the resistance near CNP. The variation
of the resistance with the gate voltage and temperature is shown in Fig. 3. The evolution of the
resistance in the samples with $d=8.5$ nm and $d=7.5$ nm resembles that for single well 2D
TIs \cite {Gusev, Olshanetsky}. For 8.5 nm sample, the resistance decreases
sharply at $T > 20$ K while saturating below 15 K. For 7.5 nm sample, a decrease in the
resistance with $T$ starts below 10 K, indicating a small mobility gap of 0.5-1 meV. Fitting the
resistance dependence on $T$ by the activation law $R \sim \exp(\Delta/2 k_B T)$, we find
the activation gap $\Delta=15$ meV for 8.5 nm DQW and $\Delta=0.5$ meV for 7.5 nm DQW.
These values agree with calculated gap energies (see Fig. 1), taking into account the $\pm 0.3$
nm uncertainty in the knowledge of $t$ and $d$. Below 10 K the conductance is saturated with
temperature, demonstrating no significant temperature dependence, which also agrees with the
resistance behavior in single well 2D TIs \cite{Gusev, suppl}. Therefore, the temperature
dependence of the resistance in 8.5 nm and 7.5 nm DQWs justifies that the bulk gap exists
so that the transport through the edge channels is expectable.

The DQWs with $d=6.5$ nm demonstrate the opposite temperature dependence: the resistance
increases with increasing $T$, indicating a metallic type of conductivity. A similar behavior
is seen \cite{suppl} in a single HgTe well with critical width $d \simeq d_c$,
which is a gapless Dirac fermion system. With increasing $T$, the resistance peak in
Fig. 3 (c) shifts to the hole side and becomes wider. This behavior is related to asymmetry
of the energy spectrum, and described within the bulk transport picture as follows. The bulk
resistance reaches a maximum when the chemical potential stays in the point where the bands
touch each other. As the density of states in the valence band is larger, for this position
of the chemical potential the hole density is larger than the electron one, and the
difference in the densities increases with $T$. Thus, the gate voltage corresponding
to the peak resistance is shifted away from the CNP to the hole side. In summary, the
temperature dependence of the resistance in 6.5 nm DQWs justifies that the gap is closed
and the transport occurs through the bulk.

\begin{figure}[ht!]
\includegraphics[width=8.5cm]{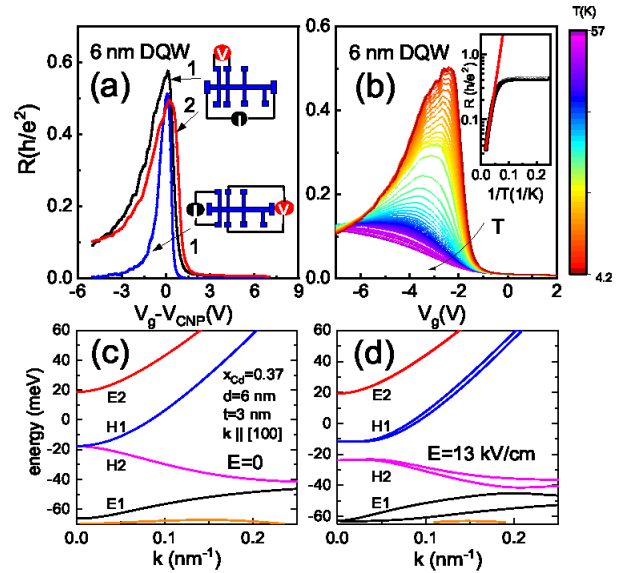}
\caption{(Color online) Experimental data and energy spectrum calculations for DQW structure
with $d=6$ nm: local and nonlocal resistances as functions of the gate voltage at $T=4.2$ K (a),
the evolution of the local resistance of sample 2 with temperature and the dependence of the
resistance at CNP on $1/T$ (b), the energy spectrum for symmetric structure (c), and the energy
spectrum in the presence of a transverse electric field $E=13$ kV/cm (d).}
\end{figure}

Finally, we present the results for DQWs with $d=6$ nm (Fig. 4), wich show a special
behavior. According to our calculations [Fig. 1 (e) and Fig. 4 (c)], this structure
is expected to be a gapless semimetal, while the transport measurements demonstrate that
it is actually a 2D TI with the activation gap $\Delta=11$ meV and a nonlocal response similar
to those in Fig. 2 (a,b). This finding can be explained by the presence of an asymmetry,
presumably related to specifics of MBE growth for this particular structure, for
example, an uncontrollable chemical doping. Calculations of the energy spectrum of
asymmetric DQWs confirm this assumption and show that the observed gap can be created
by a transverse electric field $E=13$ kV/cm, which splits the heavy hole subbands
H1 and H2 apart and also leads to Rashba spin splitting of all the subbands, Fig. 4 (d).
While in bilayer graphene the transverse electric field can be controlled by double
gates \cite{maher}, fabrication of the devices with controllable gap based on MBE-grown
HgTe DQWs requires numerous technological efforts. A forthcoming development of double
gating technology for HgTe DQWs can open a possibility for electrical control of the edge
state transport, which is important both for fundamental physics and for device applications.

In conclusion, we have demonstrated phase diversity in HgTe DQWs by observing both
a topological insulator phase and a gapless semimetal phase that resembles the
bilayer graphene. We have also detected a topological insulator phase whose existence
requires the presence of a transverse asymmetry of DQWs, possibly produced by an
internal electric field. We believe that our studies will be helpful for paving the
way to the tunable topological insulators and the multilayer topological insulators.

The financial support of this work by RFBI Grant No. 18-02-00248a.,
FAPESP (Brazil), and CNPq (Brazil) is acknowledged.

\maketitle
\section{Supplementary material:Two-dimensional topological insulator state in double HgTe quantum well}

\maketitle
\section{Sample description and magnetotransport measurements}

In the main text, we describe the experimental results for double quantum well (DQW)
samples in zero magnetic field. The structures containing HgTe quantum well layers
placed between Cd$_{x}$Hg$_{1-x}$Te barriers with surface orientation [013]
studied here were grown by molecular beam epitaxy (MBE). A schematic view of
the layer structure for the samples with 6.5 nm well width is presented in
Fig. 1 (a). As shown in the previous publications \cite{kvon1}, the use of
substrates inclined to the singular orientations may lead to growth
of more perfect films. Therefore, the efficient growth of alloys is done
predominantly on the substrates with surface orientations of [013], which deviate
from the singular orientation by approximately $19^{\circ}$. To prepare
the gate, a dielectric layer containing 200 nm SiO$_{2}$ was first grown
on the structure using the plasmochemical method. Then, the TiAu gate was deposited.

Fabrication of ohmic contact to HgTe quantum well is similar to that for other 2D systems,
such as GaAs quantum wells: the contacts were formed by the burning of indium directly to
the surface of large contact pads. Modulation-doped HgTe/CdHgTe quantum wells are typically
grown at 180$^{\rm o}$ C. Therefore, in contrast to III-V compounds, the temperatures during the
contact fabrication process are relatively low. On each contact pad, the indium diffuses
vertically down, providing ohmic contact to both quantum wells, with the contact resistance
in the range of 10-50 kOhm. During the AC measurements we always control Y-component of the
impedance that never excess 5\% of the total impedance, which demonstrates good ohmicity
of the contacts.

\begin{figure}[ht!]
\includegraphics[width=9cm,clip=]{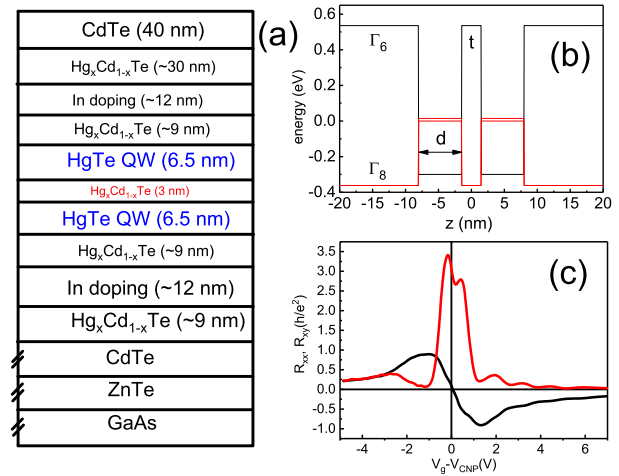}
\caption{\label{fig.1}(Color online) (a) Schematic layer structure
of a wafer containing 6.5 nm wide HgTe DQW. The corresponding band
diagram (b) and the longitudinal and Hall resistances (c) for $B=4$ T and $T=4.2$ K.}
\end{figure}

Low-temperature magnetotransport measurements were carried out to identify the
charge neutrality point (CNP) and to obtain a relation between the induced
carrier density and the gate voltage $V_{g}$. By sweeping the gate voltage, we have
found a continuous crossover from electron to hole type of conductivity, when the
Fermi level is shifted from the conduction to the valence band. The possibility to
tune the carrier density electrostatically enables the observation of electron and
hole transport in a single device, making it possible to control filling of Landau
levels in the presence of a perpendicular magnetic field. Figure 1 (c) shows the
longitudinal and Hall resistances, $R_{xx}$ and $R_{xy}$, as functions of the gate
voltage at a fixed magnetic field $B=4$ T. The plateaux with $R_{xy}=-h/\nu e^{2}$,
accompanied by the minima in $R_{xx}$, are clearly seen for electrons with filling
factor $\nu$ from $1$ to $5$. As $V_{g}$ is swept through the CNP, the longitudinal resistivity
shows a maximum, whereas $R_{xy}$ goes gradually through zero from negative on the electron
side to positive on the hole side. It is a clear indication of CNP position and
electron-hole crossover as a function of the gate voltage.
From the resistance minima, we obtain the density variation with gate voltage $0.86 \times
10^{11}$ cm$^{-2}$V$^{-1}$, which coincides with the density variation extracted from
the capacitance measurements.

The Shubnikov-de Haas oscillations was measured for all DQWs in the region of electron
conductivity. The magnetoresistance shows two sets of oscillations corresponding to
two branches of conduction-band carriers. An example of such oscillations is given in
Fig. 2.

In the quantum Hall regime, the bilayer system has a rich phase diagram in comparison
with a single well. We performed transport measurements in HgTe DQWs and observed numerous
crossings between Landau levels for both gapless and gapped samples, the results will be
reported elsewhere.

\begin{figure}[ht!]
\includegraphics[width=9cm,clip=]{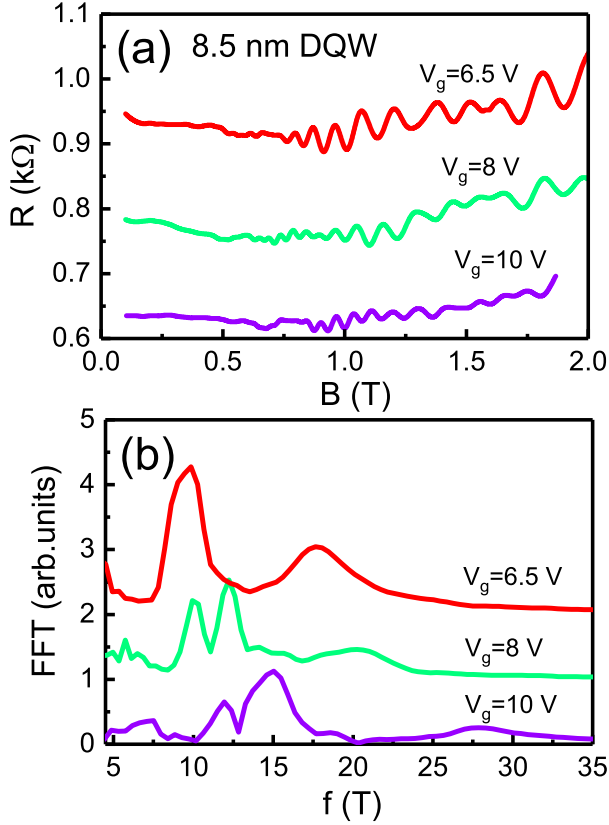}
\caption{\label{fig.2}(Color online) Shubnikov-de Haas oscillations at different gate voltages
(a) and their fast Fourier transform (b) for a DQW with $d=8.5$ nm.}
\end{figure}

\section{Transport measurements in single quantum wells}

The 2D topological insulator (TI) phase state has been observed previously in single
HgTe/Cd$_{x}$Hg$_{1-x}$Te quantum wells in the inverted band regime, when the well width
is larger than the critical width, $d>d_{c}$ \cite{Kane, Bernevig}. For $x=0.65$, one has $d_{c}=
6.2 - 6.5$ nm, depending on the surface orientation and strain. When $d=d_{c}$, the band
gap collapses, and the energy spectrum corresponds to gapless single cone Dirac fermions (DF).
In HgTe-based DQWs, modification of the spectrum depends also on the tunneling barrier
thickness and height, as has been mentioned in the main text. It is useful to compare
transport results obtained in double well and single well structures. The temperature
dependence of the resistance in DQWs for both gapless and gapped samples has been shown
in the main text. In Fig. 2 we show the variation of the resistance with gate voltage and
temperature for single quantum well with $d=6.4$ nm, which is expected to be a gapless
DF system \cite{buttner, kvon}, and for single quantum well with $d=8$ nm, which is a
2D TI \cite{Konig, Gusev}. For these two samples, the temperature behavior
is opposite: the resistance \textbf{increases} in the DF system and \textbf{decreases} in the
TI. For $T > 15$ K, the temperature dependence of the resistance in TI fits very well the
activation law $R \sim \exp(\Delta/2 k_B T)$, where $\Delta$ is the activation (mobility)
gap. The thermally activated behavior corresponds to a gap of 15 meV between the conduction
and valence bands in the HgTe well. For $T < 10$ K, the resistance is saturated with
temperature, demonstrating no significant temperature dependence. In the DF system,
the shift of the resistance peak with temperature is most likely related to
electron-hole asymmetry, as in the case of gapless DQWs. In summary, the
dependence of the resistance on temperature and gate voltage in the gapped and
gapless single quantum wells is similar to that observed in the gapped and
gapless DQWs.

\begin{figure}[ht!]
\includegraphics[width=8.7cm,clip=]{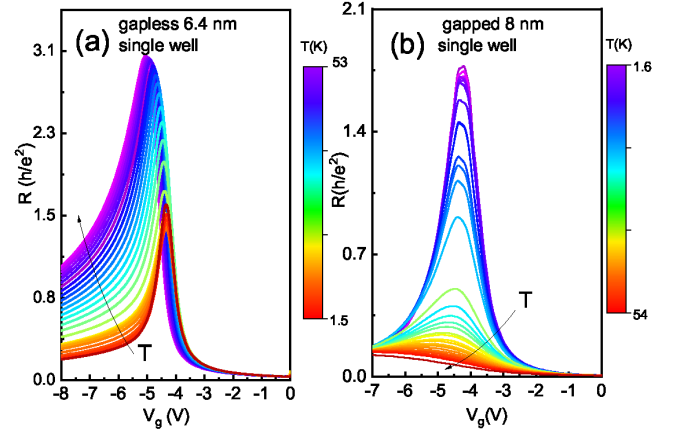}
\caption{\label{fig.3}(Color online) Resistance of the single 6.4 nm (a) and 8 nm (b)
HgTe quantum wells with Cd$_{0.65}$Hg$_{0.35}$Te barriers as a function of the gate
voltage for different temperatures.}
\end{figure}

\section{Calculations of energy spectrum}

The numerical calculations of electron energy spectrum are based on the $6 \times 6$ Kane
Hamiltonian whose description with applications for different orientations of the heterointerface
is given in Ref. \cite{raichev}. The energy is counted from the $\Gamma_8$ heavy hole band edge
in HgTe. The relative positions of the band edges of $\Gamma_6$ in HgTe, $\Gamma_6$ in CdTe, and $\Gamma_8$ in CdTe are $-0.3$ eV, $0.99$ eV, and $-0.56$ eV, respectively. The energy separation
between $\Gamma_8$ and $\Gamma_7$ bands is 1.0 eV for both HgTe and the barrier region. The Kane
matrix element $P$ and the Luttinger parameters $\gamma_{1}$, $\gamma_{2}$, and $\gamma_{3}$
for HgTe and CdTe are taken from Ref. \cite{novik}. Also, a strain-induced shift $\varepsilon_s=22$ meV
of light hole band with respect to heavy hole band in HgTe layer on CdTe substrate \cite{brune}
has been taken into account. For structures with Cd$_{x}$Hg$_{1-x}$Te barriers,
the band edge energies and Luttinger parameters in the barriers, as well as the strain-induced shift
in HgTe, have been obtained by a linear interpolation with respect to Cd content $x$. The Schroedinger
equation for columnar wave function is a $6 \times 6$ matrix
differential equation of the second order with one independent variable, the growth axis coordinate
$z$. To describe asymmetrical structures, the $z$-dependent steplike band energy profile [Fig. 1 (b)]
was modified by adding the electrostatic potential energy $eEz$. After discretization along $z$,
the differential equation is transformed to a finite-difference equation, which is solved by
using the Thomas algorithm generalized for the case of matrix coefficients.

The influence of the interface orientation on the band structure of DQWs is similar to that for
single well systems \cite{raichev}. Compared to [001]-grown structures, the [013]-grown DQWs
show shifts of the 2D subbands by several meV and modification of their dispersion, and also a
hybridization of electronlike subbands with some heavy hole subbands even at zero ${\bf k}$,
as seen in Fig. 1 (e) and (f) of the main text.

\section{Resistance of multiprobe 2D TI device with edge backscattering}

To derive Eq. (1) of the main text, the edge channels encircling the perimeter of
the 2D TI sample are represented by an equivalent circuit of resistors in series,
each individual resistor corresponds to a segment between an adjacent pair of leads $(i,j)$ with the two-terminal resistance $(h/e^2)(1+L_{ij}/l)$, where $L_{ij}$ is
the length of the segment. Applying the Kirchhoffs rules, one obtains Eq. (1),
where $V_{i,j}$ is the voltage measured between the leads $i$ and $j$ while the
current $I_{n,m}$ is maintained between the leads $n$ and $m$.

The effective lengths of the edge channel segments in the device shown in the top
of Fig. 2 of the main text are estimated as $L_{12}=4$, $L_{23}=6$, $L_{34}=12$, $L_{45}=37.2$, $L_{56}=6$, $L_{67}=40$, $L_{78}=16$, $L_{89}=10$, and $L_{91}=6$
(all the length are in $\mu$m). The estimates are obtained by adding the lengths
of vertical line segments under the gate (approximately, the top part adds 2 $\mu$m and
the bottom part adds 4 $\mu$m) to the geometrical lengths between the contacts.

\begin{figure}[ht!]
\includegraphics[width=9.5cm,clip=]{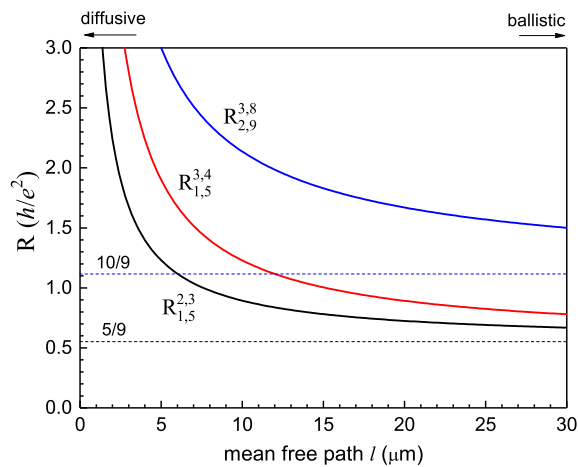}
\caption{\label{fig.4}(Color online) Calculated resistances in the local and nonlocal
measurement configurations as functions of the mean free path length $l$ characterizing
probability of backscttering in the edge channels. The horizontal dashed lines show the
resistances in the limit of ballistic transport, $l \rightarrow \infty$.}
\end{figure}

Applying Eq. (1) of the main text, one can plot the dependence of multiprobe resistances on
the mean free path length $l$. The results are shown in Fig. 4. Whereas the theoretical model
explains the observed inequality $R^{2,3}_{1,5} <  R^{3,4}_{1,5}$ in the local configurations,
the calculated nonlocal resistances $R^{3,8}_{2,9}$ appear to be considerably larger than the
observed ones. Therefore, accounting for the backscattering within the edge channels is
insufficient for explanation of the observed resistances in the 2D TI HgTe-based systems.
On the other hand, a dramatic suppression of nonlocal resistances can occur due to leakage
of electrons from the edge channel to the 2D bulk "puddles" placed nearby the edge and also
due to partial short circuiting of the edge transport by the bulk currents. Both of these
processes are expected to be important in our gapped samples because of small values of
the gap energies.


\begin{thebibliography}{25}
\bibitem{note} A different kind of gapless semimetal can be created in a single
HgTe QW by application of a strong in-plane magnetic field \cite{raichev,gusev2},
in this case the spin degeneracy is missed as well.
\end{thebibliography}
\end{document}